**Dark User Experience: From Manipulation to Deception**

**Marc Miquel-Ribé** is Lecturer in User Experience at Tecnocampus - Universitat Pompeu Fabra, Barcelona, Catalonia

# 1. Introduction

Hassenzahl (2008) defines User Experience (UX) as "the momentary feeling (good or bad) while interacting with a product or service" (p. 2). Even though this definition, or any other UX definition for that matter, do not mention that users' experiences need to be positive, the importance of experiencing positive emotions while interacting with a device is widely acknowledged. The equation "better UX = more business"[1] is the motto that determined the industry to embrace this field, and at the same time, opened the path for the UX to go beyond usability guidelines and the human factors studies. It is clear that, by facilitating tasks to the users and by addressing their needs, they will be more satisfied, more engaged, and eventually, it will have positive consequences on the business.

Nonetheless, some companies are willing to reach their economic goals at any cost, regardless of whether the customer's feels satisfied after obtaining what she needed. They wonder: how can we design the user interface in order to increase the possibilities of reaching our objectives in a much more effective way? Metrics related to purchase conversions, user retention and engagement become the main focus of design, and any UX research conducted is aimed at understanding the user's needs or preferences is used for this purpose. In these cases, design is mainly aimed at increasing the company's revenue, and companies sometimes cross the ethical line with this goal in mind. Such a phenomenon is called Dark UX.

*Dark UX applies knowledge about users in order to design for the companies' benefit, even if that implies making users do actions they did not intent to do and would not have done in other circumstances.* One of the most recognized Dark UX practices are called Dark Patterns (also known as interfaces designed to trick) and they are frequently encountered in many types of technology applications, from e-commerce websites to video games. Such interfaces are unethical because they imply negative consequences the user would have avoided, had he been informed on the matter. This is precisely where their power lies: these interfaces are effective because the user is not aware of their goal, of the way they work, or of their existence altogether.

In this chapter I identify Dark UX practices according to the way they *manipulate* and *deceive* users. I trust that this will allow us to denounce some technology derived ethical problems and help us find ways to deal with them, either helping users develop a sort of design awareness or promoting the suitable changes in the current consumer rights legislation to prevent Dark UX practices beforehand. The present chapter represents a first step towards a better understanding of the Dark UX, and a useful starting point for further discussions in the design community.

## 2. Dark User Experience: Design Not in the User's Benefit

### 2.1. What Are Dark Patterns?

Dark Patterns are defined by the website darkpatterns.org as "tricks used in websites and apps that make you buy or sign up for things that you didn't mean to".[2] The term was coined by Harry Brignull in 2010, on the wake of interaction design patterns (Tidwell 2005), defined as "general repeatable solutions to commonly-occurring usability problems"[3]. The difference is that Dark Patterns are not aimed at solving user problems, but at helping businesses increase their income, often setting an almost adversarial relationship. In fact, Dark Patterns mean more than the adaptation of the mischievous door-to-door seller into the digital era; they come with a wide-range of objectives, such as: obtain more subscriptions, help a headline go viral, make the user disclose personal information and, obviously, and increase income.

It may seem surprising that the term was only proposed after more than a decade since the web had been consolidated. Previously, there had been theoretical attempts to define an ethics for persuasive technology (Berdichevsky and Neuenschwander 1999; Fogg 2003) and even discussions about the techniques that evil interfaces employ (Conti 2008; Conti and Sobiesk 2010). Yet, the need for framing the problem was not that visible until the irruption of e-commerce websites, when all sort of money and information transactions took over the Internet. The web-based library darkpatterns.org which classifies the patterns into several categories, has rapidly become popular on the Internet and in general media.[4,5] The objective of the website is to help users understand how Dark Patterns work and warn them against the dangers of becoming victims (Brignull 2011). In some cases, publicly shaming the companies by exposing their Dark Patterns with some step-by-step screenshots, has forced them to rectify or delete the dark pattern.

While most of the Dark Patterns are usually deliberately crafted during the interface design process, it is nonetheless possible that some of them may not be purposely created to trick. Yet, when websites, apps or any digital service are running, it is difficult not to realize their positive effects for the business. As a consequence, they are not removed as they give very positive results in A/B testing, i.e. the statistical tests used to check which design gives a better performance in relation to one particular metric (Sauro and Lewis 2012).

Today, Dark Patterns are present in all sorts of environments. In video games, Zagal, Björk and Lewis (2013) have defined them as patterns "used intentionally by a game creator to cause a negative experience for players that are against their best interests and happen without their consent" (p. 3). Although the definition does not mention that they are aimed at benefiting the game publisher, this is obviously so: Dark Patterns not only push the player into spending more money, but also trick them into spending more time, and lower their social value by sending spam to their contact list, among other things. In video games, Dark Patterns design is not only inscribed in the user interface but on the rules, the mechanics, and the rest of game elements. While these sorts of Dark Patterns are harder to distinguish, they also fall under our Dark UX definition, since the player sooner or later tends to regret some played sessions, which were, in the best-case scenario, a waste of time.

**2.2. How Can We Spot a Dark UX?**

Another reason why the term Dark Patterns became popular in mainstream media is because of its different clear labels to some particularly unethical technological experiences (such as "disguised ads" or "hidden costs"). Nonetheless, despite continuous e-mails with submissions made by Internet users, the darkpatterns.org site has not increased its list from the initial 10-15 patterns. This is because, in some cases, the classification of one pattern into categories may not be straightforward or clear, since some tricks, due to their complexity, may fall at the same time into two or three categories, while in other cases, patterns sent by users were nothing but complaints of business deals or services, rather than a thorough description of how the website worked.

As the scarce previous literature (Conti and Sobiesk 2010; Zagal, Björk, and Lewis 2013; Greenberg et al. 2014) suggests, classifying Dark UX into specific categories is a challenging venture. Instead, I argue that the process of spotting Dark UX could be divided into two simple steps. First, ask how a particular dark pattern can be profitable to the company. Such profit can either be directly beneficial economically or can occur by means of other intangible benefits that eventually turn into profit. Second, in order to understand how Dark

UX function, I propose focusing on two communication concepts: *manipulation* and *deception*. I believe such concepts will help us understand how Dark UX operates and ultimately help us find strategies to counter them.

# 3. From Manipulation to Deception

## 3.1. How Do They Work?

Manipulation and deception are two classic concepts that have been long debated and investigated in connection to persuasion studies in research fields such as Social Psychology and Communication. While in these fields, researchers focus on communication in its multi-modality – where verbal cues co-occur with gestural, facial and prosodic cues -, in the digital applications such as websites, mobile apps, computer software or video games, communication always takes place through user interfaces and in the interaction with their spaces and mechanisms. In fact, the use of multiple modalities such as the audiovisual is a key characteristic of technology along with persistence, anonymity, ubiquity and personalization that endows it with a higher capacity for persuasion over human persuaders (Fogg 2003).

In this sense, interaction designers pay attention to the use of typography, words, visual representation, and interface behavior in order to ease the users' tasks, and in some cases, influence them. This is what Anderson (2011) calls "seductive interaction design". The mere fact that designers want to exert influence is not necessarily bad (such as easing the process of taking a necessary pill or of doing tedious work) – as benevolent deception can sometimes be positive to the user experience (Adar, Tan, and Teevan 2013). But in Dark UX this influence comes in the form of manipulation and deception, and it is never aimed at the user's advantage.

By manipulation it is intended a sort of social influence that aims to change the behavior or perception of others through abusive, deceptive, or underhanded tactics (Braiker 2003). Therefore, manipulation involves a manipulator and a victim, with whom these tactics become effective. Deception is sometimes included as part of the manipulation repertoire, however, it does not exploit the victim's vulnerabilities but limit to the inaccuracy of the information communicated. For this reason, I prefer discussing how each is applied to create Dark UX separately.

**a) Manipulation**

In all cases, manipulation involves knowing the psychological vulnerabilities of the victim to determine which tactics are mostly effective (Simon 1996). Some of these tactics require tailoring the communication in order to drive the user motivation into the desired direction. In his book *Evil by Design,* Nodder (2013) uses the seven capital sins as metaphorical categories to explain how to communicate or interact with the customer in order to lead them to the actions you wish.

One of the most manipulative Internet tricks based on user's interest is called *"clickbait"* (Blom and Hansen 2015). It consists of using a short headline aimed at attracting the users' attention and obtaining their clicks – e.g. sometimes using mysterious phrases like "and you will not believe what happened next". Beyond the headline hyperlink there is often a piece of information not as valuable as to fulfil the previously set expectations, and users leave the site disappointed, while having helped the website increase its number of page views. This is manipulative mainly because the same information could have been included in the initial headline, and the role of the page behind the hyperlink being that of expanding the headline with some additional information. However, in most cases, this descriptive headline would have been sufficient according to the user's interest in the topic. Other times the information which follows the "clickbait" does not deliver the promised content at all, being a practice closer to deception.

Other manipulative tactics based on motivation try to tailor the context in which the user interacts, so it is harder for them to say no to a particular action at a given moment. For instance, in online multiplayer video games, *"monetized rivalries"* is a dark pattern that exploits player competitiveness (Zagal, Björk, and Lewis 2013). In games that contain it, the player can initially enjoy the game even without spending any money until, at a given moment, the game mechanics drive to a competitiveness between players that encourages them to buy some in-game options in order to achieve an in-game status they would never reach otherwise. This is called *"pay to win"* or *"pay to cheat"*, since players who use it do not rely on their skills to win. These Dark Patterns based on motivation are usually harder to identify, since they are subtler. They smoothly create a situation for when players are faced with a choice they cannot refuse. *"Privacy suckering"*[6] is a similar dark pattern usually implemented in websites which instead of asking for money, tricks users into sharing information about themselves (even information unrelated to the action they are doing or the product they are buying). Such pattern is based on users' high interest in order to force them disclose private data.

Some other manipulative tactics are based on people's lack of accurate visual perception. In fact, most of the usability principles were created in order to avoid mistakes due to these human factors. But, once developers know how such principles work, they can deliberately turn them against the users' interests. For instance, knowing that people usually choose default options[7] may be used to make customers subscribe to predefined options and eventually pay more than they would if such an option had not been automatically selected. Likewise, there is a known usability fact that states that people do not read carefully but only scan documents (Krug 2006; Weinschenk 2011), and this is precisely at the base of the dark pattern *"trick questions"*[8]. Users are asked to answer questions which at a first glance seem to require a particular kind of answer, but in reading carefully, they are asked for a totally different one. In tricky questions, several sentences are similarly or differently phrased in order to confuse the user into assuming the opposite (see Figure 1). Some advanced versions of this dark pattern combine tricky questions with relevant information hidden in long paragraphs, instead of using a proper structure with headlines, subtitles and visual hierarchies. Perception-based Dark Patterns are designed to play on users' vulnerabilities and exploit human cognition errors in a not too obvious way. Other tactics are *"disguised ads"*[9], banners that are designed to appear exactly as the real information but which are actually paid ads; or *"misdirection"*[10], which implies hiding relevant information or options by means of size and color, while making other options and information salient to distract users from the relevant ones. Unfortunately, the possibilities to create different variations of "misdirection" patterns are immense.

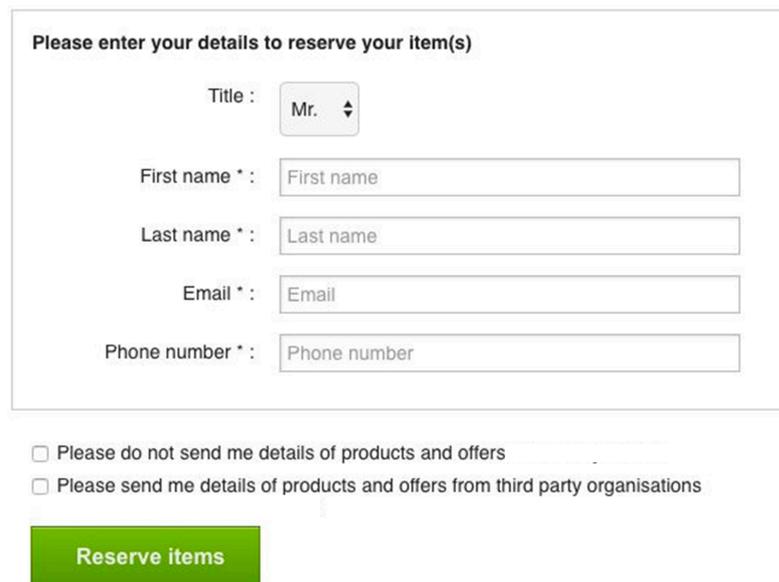

Figure 1. Trick Question used at a popular British home appliances e-commerce website[11]

A last type of manipulative tactic is based on creating frustration and disorientation so the user eventually gives up and tacitly accepts the current situation – whether it is paying a monthly fee or maintaining personal data on a server. This is a tactic commonly used in processes of unsubscription, both online and offline. While in the latter case, customers are usually asked to call a number and provide extremely detailed information in order to be able to unsubscribe, in the former case, the website's architecture and usability principles are designed to hinder the navigation. The dark pattern *"roach motel"*[12] is aimed at making it difficult to users to reach what they are looking for. All the aspects that information architecture field pays attention to, from menu to label clarity, can be perverted to create a great labyrinth where the exit that allows unsubscribing is always far. In fact, this dark pattern combines perfectly with the previously explained "misdirection", as they both take advantage of the user's lack of visual perception.

**b) Deception**

Differently from manipulation, by deception it is intended propagating information which is not true, or omitting relevant information (Buller and Burgoon 1996). Thus, deception comes with different shades of grey depending on how (un)clear communication is, and therefore, whether it allows the user to make decisions freely. For instance, *"bait and switch"*[13] consists in first showing how a part of the interface works, and later changing it without warning the user. The user is taught the meaning of a particular button, so when he sees others of the same color or shape, he is induced to infer that these new buttons will respond in a similar way to the previous ones. However, such new buttons, which share the same appearance with the previous one, accomplish a different function. For example, in Figure 2, we can see a pop-up from a popular American operating system asking the user to update. In this case, and contrary to what all users learnt from previous versions of the operating system, clicking the close button activates the software upgrade instead of closing the popup. This is totally deceptive as there is no new information alerting of this behavior change.

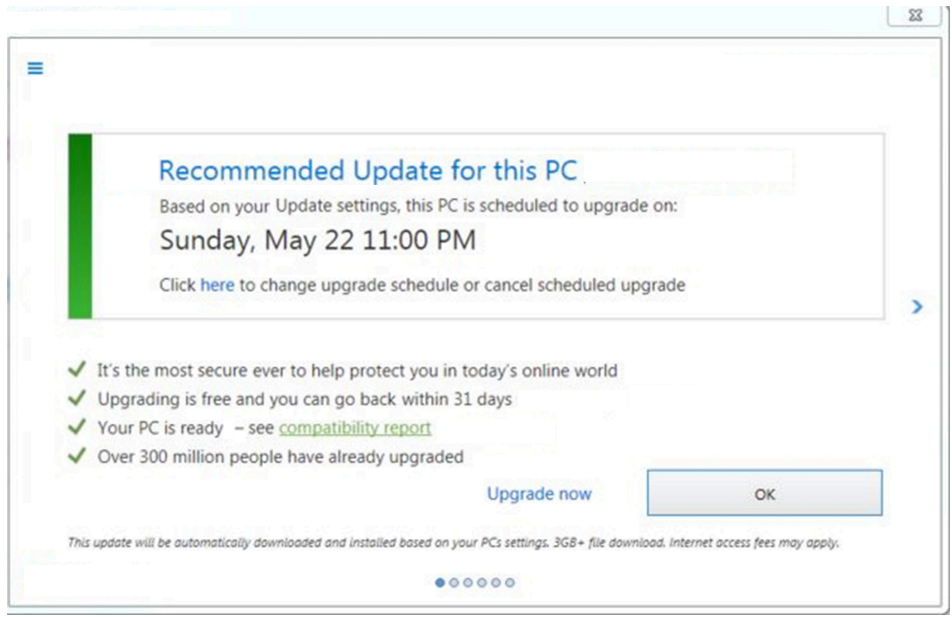

Figure 2. Popular American operating system upgrade bait and switch detected on June, 2016

Other deceptive tactics are based on not explaining or disguising the real functioning of the service, so the user assumes it works in some controllable or expected way when in fact it does not. These tactics have negative effects on game experiences. For instance, one such famous pattern is the *"near-miss"* in video gambling reel machines. In this sort of games of chance, when three drawings are coincident (e.g. cherries) the player wins a jackpot. When one of the three is distinct from the others, we say it is a near-miss. In terms of probabilities and rewards given to the player, the near-miss is nevertheless a non-win (like when all three drawings are different one from another), however, the effects of a near-miss on the player's behavior are extremely powerful. Several studies (Côté et al. 2003; Clark et al. 2009) point out that near-misses prolong the gambling, hence companies tend to display more "near-misses" than there actually are, in order to keep the player engaged and make him continue hoping and spending money. Such deceptive strategies that hide relevant information and let the users make incorrect inferences (i.e. that they are about to win) have been a controversial topic for decades.

Finally, a last type of deceptive tactics is based on hiding some relevant information at some key moments, such as for instance when decisions are being made. The Dark Patterns employing these tactics are usually directly related to spending money. For instance, the dark pattern *"hidden costs"*[14] appears in a multiple-step checkout process which shows a very attractive initial price, but which changes completely at the last step due to extra charges (whether taxes, shipping costs or "care and handling fees"). By the time the user reaches this last step, he had already invested effort in filling in all the details, that he might be willing to accept paying an extra fee rather than starting over with some other online shop. If all the information were given in advance, the user could compare prices and make informed decisions.

Another tactic that implies hiding information during critical moments is based on encouraging users to use *"credits"* or tokens during the game, instead of real money (Nodder 2013). Such a strategy is very common in games (both in mobiles and video gambling machines). In this case, the only time when the user is totally aware of the quantity of money spent is at the beginning and at the end of the game or experience. At any other moments, since the real currency is masked by a fictitious currency, the values spent during the experience do not trigger the same alert messages at a conscious level, which would otherwise moderate the spending impulse. The value of the credit can be related to an in-game narrative, and the player, completely engaged in the game, can even lose the interest for the credits more easily than he would for real money. While many games promote in-game purchases, not all of them make visible (or easy to remember) the equivalence between credits and the real currency, hence being even more deceptive.

## 4. Conclusions

Dark UX applies knowledge about the user for the company's benefit, even pushing the user to do actions he might not have intended to do and would have certainly not done in other circumstances. It is called 'dark' not only because it is unethical in its tactics and outcomes, but because the user is usually unaware while it is happening. The Dark Patterns are a popular way to identify and expose Dark UX, and there have been several attempts to classify them into particular categories (Zagal, Björk, and Lewis 2013; Conti and Sobiesk 2010). Considering that sometimes this classification process may be unclear, I proposed distinguishing between tactics based on manipulation and others based on deception, as a useful way to reflect on how each dark pattern operates.

Manipulation implies taking advantage of the users' psychological weaknesses and it can relate to motivation and perception. Many Dark Patterns are based on usability facts and principles aimed at improving the user experience, and which are now applied in the exact opposite direction with respect to their initial objective.

Deception works instead by clearly changing or hiding relevant information in order to alter the decision-making processes. They are not as subtle as the manipulative ones, and they can be identified in a clearer way. Nonetheless, some sites are more original and include Dark Patterns which are a combination of the two.

I argue that the benefit of the suggested distinction (manipulation and deception) is two-fold. First, people may better understand how Dark Patterns operate and can avoid being tricked by them. As I already mentioned elsewhere (Miquel-Ribé 2014; Miquel-Ribé 2015), the manipulation-based Dark Patterns can be countered with a better education or more *design awareness*. As far as deception-based Dark Patterns are concerned (like "near misses" or "hidden costs"), since they are easier to objectify, they are probably easier to fight with legislation, and some had already been banned in the European Union[15].

Second, the few approaches to study Dark UX often lack theory and empirical investigation. In order to classify Dark UX, it is important to really understand their nature, their components and how they are processed by people. Hence, key concepts such as manipulation and deception, can directly relate to psychological limitations and information qualities, which allow operationalization and measurement of the Dark UX practices elements. Empirical research based on these concepts might inform users, researchers and legislators on which practices are more deceptive or manipulative, and at the same time, which eventually will provide a stronger knowledge base to demand technological applications free from them.

## 5. Endnotes

# End notes

[1] https://www.forbes.com/sites/forbesagencycouncil/2017/03/23/the-bottom-line-why-good-ux-design-means-better-business
[2] http://darkpatterns.org
[3] https://www.interaction-design.org/literature/book/the-glossary-of-human-computer-interaction/interaction-design-patterns
[4] https://www.theverge.com/2013/8/29/4640308/dark-patterns-inside-the-interfaces-designed-to-trick-you
[5] https://www.wired.com/2014/02/dark-patterns-user-interfaces-designed-trick-people/
[6] https://darkpatterns.org/types-of-dark-pattern/privacy-zuckering
[7] https://www.nngroup.com/articles/the-power-of-defaults/
[8] https://darkpatterns.org/types-of-dark-pattern/trick-questions
[9] https://darkpatterns.org/types-of-dark-pattern/disguised-ads
[10] https://darkpatterns.org/types-of-dark-pattern/misdirection
[11] https://darkpatterns.org/types-of-dark-pattern/trick-questions
[12] https://darkpatterns.org/types-of-dark-pattern/roach-motel
[13] https://darkpatterns.org/types-of-dark-pattern/bait-and-switch
[14] https://darkpatterns.org/types-of-dark-pattern/hidden-costs
[15] https://www.boe.es/doue/2011/304/L00064-00088.pdf